\documentstyle[prl,aps,multicol,epsf]{revtex}
\begin{document}
\title{Current noise in a irradiated point contact}
\author{Y. Levinson }
\address{ Department of Condensed Matter Physics, The Weizmann 
Institute of Science,76100 Rehovot, Israel}
\author { P. W\"olfle}
\address{Institut f\"ur Theorie der Kondensierten Materie,
Universit\"at Karlsruhe, 76128 Karlsruhe, Germany}
\date{24 January 1999}
\maketitle
\begin {abstract}
We propose a new approach to calculate current and current correlations
in a ballistic quantum point contact 
interacting with a classical field.
The approach is based on the concept of scattering states
for a {\it time dependent} Hamiltonian neglecting electron-electron
interaction.
Using this approach we calculated the spectra of the current noise
in a biased point contact irradiated by a weak random field.
For typical radiation frequencies $\nu$ less than the temperature $T$
and the bias voltage $V$ we find a narrow peak of width $\nu$
on top of a broad background of width $\max (T,eV)$.
\end {abstract}
\pacs {PACS numbers: }
\vspace{-1 cm}
\begin{multicols}{2}
Current fluctuations in ballistic quantum point contacts (PC)
attract now  much  attention,
both in theory \cite{Khl87,Les,But90,YuKo,EY,But,MarLan92}
and experiment \cite{Rez95,Kum96,BirJonSch95}
partly because of the assumed possibility to measure
fractional charges in shot-noise \cite{Rez}
and to probe other non Fermi liquid properties\cite{Ma}.
We are interested in current fluctuations
in a ballistic PC biased by applied voltage 
and irradiated by external field.
{\it Time averaged} current in microstructures
under such conditions (photon-assisted current) was 
investigated experimentally in point contacts \cite{Wys93,Jan94} 
and quantum dots \cite{Kou94,Bli95,FujTar97,Oos97}.

Current in a PC irradiated
by a {\it monochromatic} a.c. field 
was discussed in \cite{DatAna,Gri,Pas82,ChuTan,YeyFlo91}. 
Current fluctuations under such an irradiation
where discussed in \cite{IvLev,LesLev}.

We consider a classical field which can be  coherent 
(e.g. microwave radiation) or incoherent 
(e.g. representing the environment at high enough
temperature or a heat phonon pulse).
We assume that the field does not irradiate the leads between 
which the bias is applied. [Which means e.g. modulated gate voltage,
not modulated bias voltage.] 
We model this situation considering
a 1D channel with a time-dependent barrier potential $U(x,t)$.
The d.c. part of the potential $U_{0}(x)$ is due to the 
squeezing of the PC while the a.c. part $\delta U(x,t)$
is due to the field. 

Consider the 1D Schr\"odinger equation
$i{(\partial/\partial t)\psi}=H\psi $ for one
particle  with a time dependent Hamiltonian 
$H=- \nabla^2/2m + U(x,t) $,             
where the barrier potential $U(x,t)=0$  at $x\rightarrow \pm\infty$
for all $t.$
For any energy  $\epsilon_k \equiv k^{2}/2m>0$ (with $k>0$)
we define {\it time dependent} scattering states 
 $\chi_k^\sigma (x,t),\; \sigma = \pm$, as solutions of the
Schr\"odinger equation with the following boundary conditions.
At $x\rightarrow -\infty$
\begin{eqnarray}
&&\chi_{k}^+(x,t)=L^{-1/2}\Big[e^{-i\epsilon_{k} t+ikx}+
\sum_{k'}r_{kk'}e^{-i\epsilon_{k'} t-ik'x}\Big],\\ \nonumber
&&\chi_{k}^-(x,t) = L^{-1/2}\sum_{k'}{\tilde t}_{kk'}
e^{-i\epsilon_{k'} t-ik'x},
\label{bcm} 
\end{eqnarray}
and at $x\rightarrow +\infty $
\begin{eqnarray}
\label{bcp}
&&\chi_{k}^-(x,t)=L^{-1/2}\left[e^{-i\epsilon_{k} t-ikx}+
\sum_{k'}{\tilde r}_{kk'}e^{-i\epsilon_{k'} t+ik'x}\right],\\ \nonumber
&& \chi_{k}^+(x,t)=L^{-1/2}\sum_{k'}t_{kk'}
e^{-i\epsilon_{k'} t+ik'x},
\end{eqnarray}  
where $L$ is the normalization length. 
One obtains time independent reflection and transmission coefficients
$r_{kk'}$ and $t_{kk'}$
 for inelastic scattering since $U(x)=0$ for $x\rightarrow \pm \infty$ 
and any outgoing wave can be presented as a superposition
of {\it time dependent} plane waves with {\it time independent}
coefficients.
For any fixed time $t$, 
the states $\chi_k^\sigma(x,t)$ form an
orthonomal and complete basis  since the states
 $\chi_{k}^{\pm}$ can be obtained from a complete set
of the functions $L^{-1/2}e^{\pm ikx}$
by the unitary transformation corresponding to time evolution
with Hamiltonian $H(x,t)$ from $t=-\infty$.
The solutions are labeled according to the energy of the incoming wave.
The sums over $k'$ represent inelastic scattering
in transmission and reflection by the a.c. barrier.
They are restricted to a  $k'$ interval defined 
by $|\epsilon_{k'}-\epsilon_{k}|\alt \nu$, where $\nu$
is the typical frequency of the barrier variation.
For harmonic oscillations of the barrier these  scattering states
reduce to those used in \cite{IvLev,LesLev,Bor97}.

For a d.c. barrier the reflection and transmission coefficients are
diagonal 
$r_{kk'}=r_{k}\delta_{kk'},{\tilde r}_{kk'}={\tilde r}_{k}\delta_{kk'},
t_{kk'}={\tilde t}_{kk'}=t_{k}\delta_{kk'}$
and the time dependent scattering states 
are reduced to the usual ones, i.e.
$\chi_{k}^\sigma (x,t)= e^{-i\epsilon_{k} t}\chi_{k}^\sigma(x)$. 

The many-particle wave functions of the system, neglecting
electron-electron interactions are given by Slater determinants 
of the
scattering
states $\chi_k^\sigma(x,t)$. We define creation (annihilation)
operators
$a_{k\sigma}^+(a_{k\sigma})$ for electrons in scattering
 states
$\chi_k^\sigma(x,t)$.  Employing the usual interpretation of the
scattering states as describing the transition amplitudes for a
wavepacket approaching the interaction zone from the far left side
$(\chi_k^+)$ or far right side $(\chi_k^-)$ to be transmitted or
reflected, we may interprete the operators $a_{k+}^+$ and $a_{k-}^+$
as creating electrons in the left and right leads $a$ and $b$, 
respectively.
Assuming the electron system in the leads to be 
modelled by 
a free
Fermi gas in equilibrium at temperature $T$ and chemical
potentials $\mu_a$ and $\mu_b$, the average occupation numbers of the states
$\chi_k^\sigma$ are given by 
$\langle a_{k\sigma}^+a_{k'\sigma'}\rangle =
\delta_{kk'}\delta_{\sigma\sigma{'}}n_{k\sigma}$, with
$n_{k\pm}$ being  
the respective Fermi functions.  
The bias
voltage $V$ applied to the point contact between the leads $a$ and $b$ 
is  $eV = - (\mu_a - \mu_b)$ with $ e > 0$.

The time-dependent electron field operator can then be represented in
the following way
\begin{equation}
\Psi(x,t) = \sum_{k\sigma} a_{k\sigma} \chi_k^\sigma (x,t).
\label{fo}
\end{equation}
  The representation Eq.(\ref{fo}) 
assumes that the leads are ``black bodies'' and do not reflect
incoming electron waves \cite{Lan92}. 
Our presentation of the electron field can be considered
as a formalization of the wave-packet
approach used in \cite{MarLan92} and is similar to that used in
\cite{But} for a d.c. barrier.  

The {\it time-dependent} current operator is
\begin{eqnarray} 
j(x,t)=(ie/2m) \Psi(x,t)^{+} \nabla \Psi(x,t) + h.c. .             
\label{jop}
\end{eqnarray}
Introducing here the field operator from Eq.(\ref{fo}) we have
\begin{equation}
 j(x,t)=
\sum_{kk'\sigma\sigma{'}}  a_{k{'}\sigma{'}}^+ a_{k\sigma} 
A_{k'k}^{\sigma{'}\sigma}(x,t),
\label{jopk} 
\end{equation}
where the partial currents  are bilinear combinations of scattering states
and depend on $x,t$
\begin{equation} 
 A_{k'k}^{\sigma{'}\sigma} (x,t) =
(ie/ 2m)(\chi_{k'}^{\sigma{'}*}\nabla\chi_{k}^\sigma 
 - \chi_{k}^\sigma\nabla\chi_{k'}^{\sigma{'}{*}}) = (A_{kk'}^{\sigma\sigma{'}})^*
\label{ABC}           
\end{equation}

Using the properties of the Fermi operators one can perform the
quantum averaging and  obtain an average current
(which is space and time dependent)
\begin{eqnarray} 
\langle j(x,t)\rangle=\sum_{k,\sigma}n_{k\sigma}A_{kk}^{\sigma\sigma}(x,t).
\label{jav}
\end{eqnarray}
For a d.c. barrier one can see from Eq.(\ref{ABC})
 that the diagonal combinations
 $A_{kk}^{\sigma\sigma}$  and hence the current Eq.(\ref{jav}) 
do not depend on $x$ and $t$.

In what follows we will concentrate on the case when the
time dependent part of the 
barrier $\delta U(x,t)$ is a stationary random function of $t$,
defined by a correlator $\overline{\delta U(x,t)\delta U(x',t')}$,
(assuming  $\overline{\delta U(x,t)}=0$),
where $\overline{(...)}$ means statistical averaging.

For a randomly fluctuating barrier the current given by Eq.(\ref{jav})
has to be also statistically averaged, giving
\begin{eqnarray}
\overline{\langle j\rangle}= 
\sum_{k\sigma}n_{k\sigma}\overline{A_{kk}^{\sigma\sigma}}.          
\label{javav}
\end{eqnarray}

In the stationary case this current does not depend on $t$
and hence on $x$. Note that for an asymmetric barrier and/or
asymmetric irradiation $\overline{\langle j\rangle}\neq 0$ for $V=0$,
in general.
 
The current correlator is defined using full averaging
\begin{eqnarray}
 K(1,2) =
{1\over 2} \overline{\langle j(1)j(2)+j(2)j(1)\rangle}-
\overline{\langle j(1)\rangle}\,\,\overline{\langle j(2)\rangle},            
\label{corr}
\end{eqnarray} 
where the short notation means $1\equiv x_{1},t_{1}$
and $2\equiv x_{2},t_{2}$.
It is convenient to represent $K=K_{q}+K_{s}$, 
where the first term is the statistically averaged
quantum-mechanical  correlator 
\begin{eqnarray}
K_{q}(1,2)=
\overline{{1\over 2}  \langle j(1)j(2)+j(2)j(1)\rangle-
\langle j(1)\rangle\langle j(2)\rangle},                    
\label{corrq}
\end{eqnarray} 
while the second term is the statistical correlator
of the  quantum-mechanical current,

\begin{eqnarray}
K_{s}(1,2)=\overline{\langle j(1)\rangle\langle j(2)\rangle}-
\overline{\langle j(1)\rangle}\,\,\overline{\langle j(2)\rangle}.
\label{corrs}
\end{eqnarray}   
Note that $K_{q}=0$ if the current is classical, while
$K_{s}=0$ for a d.c. barrier.

Introducing in Eq.(\ref{corrq}) and Eq.(\ref{corrs})
the current operator from Eq.(\ref{jopk}) one finds

\begin{eqnarray}
\label{corrq1}
K_q(1,2)& =& \frac{1}{2}
\sum_{kk'\sigma\sigma{'}}\Big[n_{k\sigma}(1-n_{k'\sigma'})\overline{A_{k'k}^{\sigma'\sigma}(1)^*A_{k'k}^{\sigma'\sigma}(2)}
+ c.c.\Big]\nonumber\\
K_{s}(1,2)&=& 
\sum_{kk'\sigma\sigma'} n_{k\sigma}n_{k'\sigma'}
\overline{\delta A_{kk}^{\sigma\sigma}(1)\delta
A_{k'k'}^{\sigma'\sigma'}(2)},
\end{eqnarray} 
 where
$\delta A_{kk}^{\sigma\sigma} = A_{kk}^{\sigma\sigma} - 
\overline{A_{kk}^{\sigma\sigma}}$.
For a stationary random barrier both correlators $K_{q}$
and $K_{s}$ depend on $t_{1}-t_{2}$ and on $x_{1}, x_{2}$.

We will be interested in low-frequency "quasistationary" 
current fluctuations for which the correlator does not depend on 
$x_{1}, x_{2}$. To understand when this situation
occurs consider first a d.c. barrier.
We assume the barrier is ``simple'', i.e.
its  height is  of the order of Fermi energy
 $\epsilon_{F}$ and its length  $d$ is of the order
of the Fermi wave length $2\pi/k_{F}$
and there are no other energy or length scales,
as e.g. in a double barrier potential.
In this simple case the energy scale for $t_{k}, r_{k},\tilde{r}_{k}$
is $\epsilon_{F}$.  The scale for
$A^{\sigma\sigma{'}}$ is the same as  if these quantities are calculated
for $x\alt d$. At $x\gg d$ a new smaller scale appears.
To see it we calculate $A^{\sigma\sigma{'}}$ 
using the Eq.(\ref{bcp}) for the scattering states,
 giving at $x\gg d$, for example

\begin{eqnarray}
\label{Ckk'}
A_{k'k}^{+-}=-(e/ 2mL)e^{i(\epsilon_{k'}-\epsilon_{k})t}
\cdot \\ \nonumber
t_{k'}^{*}\left[(k+k')\tilde{r}_{k}e^{-i(k'-k)x}
-(k-k')e^{+i(k'+k)x}
\right].
\end{eqnarray}  
As we will see later the relevant momenta $k,k'$
correspond to energies $\epsilon_{k},\epsilon_{k'}$ within 
the exchange window between the Fermi distributions
in both leads $|\epsilon_{k}-\epsilon_{F}|\alt \max (eV,T)$
which is assumed to be narrow compared to $\epsilon_{F}$.
Hence for a simple barrier
one can put $k=k'=k_{F}$ everywhere except in the exponentials
(since $x$ and $t$ can be large).
As a result the fast oscillating exponentials (of Friedel type)
$e^{\pm i(k'+k)x}$ disappear.
The slow oscillating exponentials $e^{\pm i(k'-k)x}$
introduce a new energy scale  $v_{F}/x$ which is 
smaller than $\epsilon_{F}$ if $x\gg d$. 
This scale corresponds to  the inverse time of
flight from the barrier to the point where current correlations
are measured.

The correlations are quasistationary if $(k-k')x\ll 1$.
Since the relevant energy transfers are 
$\epsilon_{k'}-\epsilon_{k}\simeq \omega$,
where $\omega$ is the current fluctuation frequency,
 the relevant $k-k'\simeq\omega/v_{F}$.
If $\omega\ll \epsilon_{F}$ we can satisfy the quasistationarity
condition choosing $x$  in the interval $d\ll x\ll v_{F}/\omega$.
As a result we see that if the current correlations at frequencies
$\omega$ are measured not too far from the barrier,
at $ x\ll v_{F}/\omega$, the current fluctuations are
quasistationary and are the same in all crossections of the PC.

With these assumptions one can put $k=k'=k_{F}$ everywhere
except in the {\it time} exponentials yielding the much simpler
expressions 
\begin{equation}
\label{AAF}
A_{k'k}^{\sigma'\sigma}(1)^* A_{k'k}^{\sigma'\sigma}(2) =
\Omega_{k'k}(t_1-t_2)
A_F^{\sigma'\sigma}
\label{}
\end{equation}
where $\Omega_{k'k}(t) = (ev_F/L)^2\exp[i(\epsilon_{k'} -
\epsilon_k)t],\; A_F^{\sigma\sigma} = \mid t_F\mid^4$,
$A_F^{-\sigma\sigma} = \mid t_F\mid^2\mid r_F\mid^2$, and
$t_F, r_F$ are the 
transmission and reflection amplitudes $t_k, r_k$ at $\epsilon_k =
\epsilon_F$.     
Using this result in Eq.(\ref{corrq1}) one obtains the
current correlator for a static barrier,  calculated in \cite{Les}, 
which we 
quote in the  time domain for later comparison:
\begin{equation}
K^{(0)}(t)=(e^2/4\pi ^2)F(t)
[|t_{F}|^2+|t_{F}|^2|r_{F}|^2 (\cos (eVt)-1)],
\label{K0}
\end{equation}
where
\begin{equation}
F(t)=2\int d\epsilon\int d\epsilon 'n(\epsilon)[1-n(\epsilon)]
\cos[(\epsilon-\epsilon{'})t].
\label{F}
\end{equation}
One can see from Eq.(\ref{K0}) that $F(t)$ is (up to a factor)
the correlator of equilibrium noise in a non biased PC with $V=0$.
 
In  the case of a time-dependent barrier fluctuating with frequencies $\nu$
the exchange window is  $\max(eV,T,\nu)$.
In case the barrier fluctuations are slow, i.e  $\nu\ll \epsilon_{F}$,
the quasistationarity conditions of the current fluctuations
are the same as for a d.c. barrier. (For fast barrier fluctuations
these conditions can not be satisfied.)

In what follows we  consider the case when the radiation field is weak,
which means in our model that the fluctuating part of the barrier 
is small compared to the d.c. part, i.e. $\delta U(x,t)\ll \epsilon_{F}$.
In the lowest order the 
field induced current noise  is proportional to $\delta U ^{2}$.

The time-dependent scattering states can be expanded in powers of
$\delta U$ as
$\chi_{k}^\sigma (x,t) =e^{-i\epsilon_{k} t}
\Big[\chi_{k}^\sigma (x)+\chi_{k}^{\sigma(1)}(x,t)+
\chi_{k} ^{\sigma(2)}(x,t)+...].$
Here  $\chi_k^\sigma(x)$  
are  the scattering states for the Hamiltonian $H_{0}$
with the average barrier
$U_{0}$, while  $\chi^{(1)}, \chi^{(2)}$ 
 are slow functions of $t$.
These functions contain only outgoing waves and can be calculated
in the Born approximation using the retarded Green function
 given by \cite{MF}
\begin{eqnarray}                  
G(x,x',t)={1\over 2\pi}\int d\epsilon_{k} e^{-i\epsilon_{k} t}
{mL\over ikt_{k}}\chi_{k}^+(x_{>})\chi_{k}^-(x_{<}),
\label{Gxt1}
\end{eqnarray}
where
 $x_{>}$ and $x_{<}$ are the larger and smaller of $x$ and $x'$. 

Using the above expansion 
one can expand the averages entering
Eq.(\ref{corrq1}) and find after lengthy but 
straightforward calculation 
 the field induced parts of these averages.
Having in mind quasistationary frequencies 
and a simple barrier one obtains  using the properties of 
the scattering states for a d.c. barrier
\begin{eqnarray}
\left[
\overline{A_{k'k}^{\sigma'\sigma}(1)^{*}A_{k'k}^{\sigma'\sigma}
(2)}\right]^{(2)}=
\Omega_{k'k}(t_1-t_2)\Psi_{\sigma{'}\sigma}(t_1-t_2),
\nonumber\\
\left[\overline{\delta A_{kk}^{\sigma\sigma}(1)\delta
A_{k'k'}^{\sigma'\sigma'}
(2)}\right]^{(2)} =
(ev/L)^2\Psi_{\sigma\sigma '}(t_1-t_2),
 \label{dABC} 
\end{eqnarray}
where
\begin{equation}
\Psi_{\sigma'\sigma}(t)=(L/v_{F})^2\overline{\delta
U^{\sigma'\sigma}(t)^{*}
\delta U^{\sigma'\sigma}(0)},
\end{equation}
and the effective matrix elements are
\begin{eqnarray}
\label{me}
\delta U^{\sigma\sigma}(t) = \int dx \delta U(x,t)
[r_Ft_F^*\chi_F^-(x)\chi_F^+(x)^* -c.c],
\nonumber\\
\delta U^{-\sigma\sigma}(t) = \int dx \delta U(x,t)
[r_Ft_F^* (\mid \chi_F^+(x)\mid^2 -
\nonumber\\
\mid\chi_F^-(x)\mid^2)
+ \chi_F^-(x)^*\chi_F^+(x)(\mid t_F\mid^2 - \mid r_F\mid^2)].
\end{eqnarray}
For later use we define $\Psi(t) \equiv \Psi_{\sigma\sigma}(t) =
\Psi^*(t)$ and $\Phi(t) = \Psi_{-\sigma\sigma}(t) = \Phi(-t)^*$.

Now we introduce Eqs.(\ref{dABC}) into Eqs.(\ref{corrq1}) 
and find the field induced correlator, 
\begin{eqnarray}
K_{q}^{(2)}(t)&=&{1\over 2}
\sum_{kk'\sigma\sigma'} 
\Big[\Omega_{k'k}(t)
n_{k\sigma}(1-n_{k'\sigma'})\Psi_{\sigma'\sigma}(t) + c.c.\Big]
\nonumber\\
K_{s}^{(2)}(t)&=&\left({ev_{F}\over L}\right)^2
\left[\sum_{k}  (n_{k+} - n_{k-})\right]^2
\Psi(t).
\end{eqnarray}
To simplify the expressions we
 replace $\sum_{k}$ by $ (L/2\pi v_{F})\int d\epsilon$ and
 shift the arguments in the distribution functions.
Using Eq.(\ref{F}) we find
\begin{eqnarray}
\label{Kq}
K_{q}^{(2)}(t)=(e^2/8\pi ^2)F(t)[\Phi(t)e^{ieVt}+\Psi(t)+c.c.]
\end{eqnarray}
and
\begin{eqnarray}
\label{Ks}
K_{s}^{(2)}(t)=(e^2/8\pi ^2)(eV)^2\Psi(t).
\end{eqnarray}

It is obvious from the above derivation that the results Eqs.(\ref{Kq}),
(\ref{Ks}) and (\ref{K0}) are valid only for $t\gg \epsilon_{F}^{-1}$,
i.e. for Fourier components $\omega\ll \epsilon_{F}$.
The spectra $\Phi(\omega)$ and $\Psi(\omega)$ of the
functions $\Phi(t)$ and $\Psi(t)$ contain only such
frequencies, but this is not the case for $F(t)$
which has a singularity $t^{-2}$ at $t\rightarrow 0$.
The Fourier transform of this function can be presented as
$F(\omega)=
2|\omega| \left[{\cal N}(|\omega|)+{1\over 2}\right]$,
where
${\cal N}(\omega)=\left[\exp(\omega/T)-1 \right]^{-1}$
is the Planck distribution. Because of the zero-point fluctuations
$F(\omega)$ has no natural cutoff below $\epsilon_{F}$.
(A cutoff at $\epsilon_{F}$ exists because of fast oscillations of
the partial currents Eq.(\ref{ABC}) when 
$|\epsilon_{k}-\epsilon_{k'}|\agt \epsilon_{F}$.)

The zero-point fluctuations disappear if one calculates
the ``shot noise'' contributions, i.e. subtracts from all correlators
the values at $V=0$.
For a static PC the shot noise is
\begin{eqnarray}
\label{K0V}
K^{(0)}_{V}(t)\equiv  K^{(0)}(t)-K^{(0)}(t)|_{V=0}=\\ \nonumber
(e^2/4\pi ^2)|t_{F}|^2|r_{F}|^2 F(t)(\cos eVt -1),
\end{eqnarray}
while the  shot-noise induced by a time-dependent field is given by 
\begin{eqnarray}
\label{K2V}
K^{(2)}_{V}(t)\equiv K^{(2)}(t)-K^{(2)}(t)|_{V=0}=\\ \nonumber
(e^2/8\pi ^2)[F(t)(e^{ieVt}-1)\Phi(t)+(eV)^2\Psi(t)+c.c.].
\end{eqnarray}
In Eq.(\ref{K0V}) the singularity of $F(t)$ is compensated
by the factor $(\cos eVt-1)$. The same happens in
Eq.(\ref{K2V}) since $\Phi(t)$ is real at small $t$.

The spectra $S^{(2)}_{V}(\omega)$ of the correlator $K^{(2)}_{V}(t)$
contain two different contributions. One, due to $K_{q}$,
is a convolution of $S^{(0)}_{V}(\omega)$, the shot noise spectrum
in a static PC, with $\Phi(\omega)$.
The presence of such a contribution which  containes
frequencies $\omega\simeq T\pm eV\pm \nu$, is almost obvious.
The less obvious result is the second contribution due to $K_{s}$,
which is proportional to $\Psi(\omega)$ and contains only
frequecies $\nu$ of the radiation field. This contribution can be easily
separated from the first one, 
since it is temperature independent and proportional
to $V^2$. This contribution allows for a spectral resolution
of  a narrow-band radiation by a device having broad band
noise.

To understand more about the nature of the field induced current
fluctuations consider some specific cases.

First consider a symmetric barrier $U_{0}(-x)=U_{0}(x)$ exposed 
to symmetric irradiation $\delta U(-x,t)=\delta U(x,t)$.
Using $\chi^{-}_{k}(x)=\chi^{+}_{k}(-x)$ we find from 
Eqs.(\ref{Kq}), (\ref{Ks})
\begin{eqnarray}
K_{q}^{(2)}(t)=(e^2/4\pi^2)F(t)\Xi(t)\cdot \\ \nonumber
\{(|t_{F}|^2-|r_{F}|^2)^2\cos eVt +4|t_{F}|^2|r_{F}|^2\},
\end{eqnarray}
\begin{eqnarray}
K_{s}^{(2)}(t)=(e^2/4\pi^2)(eV)^2|t_{F}|^2|r_{F}|^2\Xi(t).
\end{eqnarray}
Here
$
 \Xi(t)=(L/v_{F})^2 \overline{g(t)g(0)}
$
with
$
g(t)=\int dx \delta U(x,t)\chi^{+}_{F}(x)\chi^{+}_{F}(-x)^{*}=g(t)^{*}.
$
For $V=0$ one finds the total correlator to be
\begin{eqnarray}
K^{(2)}(t)=(e^2/4\pi^2)F(t)\Xi(t).
\end{eqnarray}
It means that symmetric irradiation in a non biased PC
creates nonequilibrium current noise (even in a symmetric PC
where average current  $\overline{\langle j\rangle}$
is not created),
while applying symmetrically a voltage to both leads 
leaves the PC in equilibrium and no additional noise is created.

As a second example consider a PC exposed to irradiation 
localized to the left of the barrier at $|x|\gg d$
assuming $\delta U(x,t)$ to be a smooth function of $x$.
In this case one can use the Eqs.(\ref{bcm}) calculating the matrix elements
Eq.(\ref{me})  
and neglect terms containing Friedel oscillations.
As a result
\begin{eqnarray}
K^{(2)}(t)=(e^2/4\pi^2)F(t)|t_{F}|^2|r_{F}|^2\Xi(t)\cos eVt.
\end{eqnarray}
Taking here $V=0$ we can compare the nonequilibrium noise
in a asymmetrically irradiated non biased PC with the 
nonequilibrium noise
in a biased non irradiated PC given by Eq.(\ref{K0V}).
One can see that ``one side excitation'' of the PC
does not simulate a bias voltage.
Even if we choose the excitation to be
quasimonochromatic with frequency $\nu=eV$, in which case
$\Xi(t)\sim \cos eVt$, the nonequilibrium noise due to
irradiation contains the field amplitude.

Both considered examples demonstrate that the nonequilibrium noise
excited by irradiation differs essentially from
nonequilibrium noise excited by bias.

This work was supported by the Alexander von Humboldt Foundation,
Israel Academy of Sciences
(Y.L), the German-Israeli Foundation and the Deutsche 
Forschungsgemeinschaft (P.W).

\end{multicols}

\vspace{-0.5 cm}


\begin{references}
\vspace{-1.3cm}

\bibitem{Khl87}
V.A.Khlus, 
Sov.Phys.JETP {\bf 66} 1243 (1987)

\bibitem{Les} 
G.B.Lesovik, 
JETP Lett. {\bf 49}, 592 (1989)  

\bibitem{But90}
M.Buttiker,
Phys.Rev.Lett. {\bf 65}, 2901 (1990)

\bibitem{YuKo}
B.Yurke and G.P.Kochanski,
Phys.Rev. {\bf B41}, 8184 (1990)

\bibitem{EY}
S.R. Eric Yang,
Solid State Comm.{\bf 81}, 375 (1992)  

\bibitem{But} 
M.Buttiker, 
Phys.Rev. {\bf B46}, 12485 (1992)

\bibitem{MarLan92}
Th.Martin and R.Landauer
Phys.Rev. {\bf B45}, 1742 (1992)

\bibitem{Rez95}
M.Reznikov at al,
Phys.Rev.Lett. {\bf 75}, 3340 (1995)

\bibitem{Kum96}
A.Kumar et al,
Phys.Rev.Lett. {\bf 76}, 2778 (1996)

\bibitem{BirJonSch95}
H.Birk, M.J.M. de Jong and C.Sch\"onenberger,
Phys.Rev.Lett. {\bf 75}, 1610 (1995)

\bibitem{Rez}
M.Reznikov et al,
Supperlatt. and Microstr. {\bf 23}, 901 (1998)

\bibitem{Ma}
Th.Martin,
Supperlatt. and Microstr. {\bf 23}, 859 (1998)

\bibitem{Wys93}
R.A.Wyss et al,
Appl.Phys.Lett. {\bf 63}, 1522 (1993)

\bibitem{Jan94}
T.J.B.M.Janssen et al,
J.Phys.: Condens.Matter {\bf 6}, L163 (1994)

\bibitem{Kou94}
L.P.Kouwenhoven et al,
Phys.Rev.Lett. {\bf 73}, 3443 (1994)

\bibitem{Bli95}
R.H.Blick et al,
Appl.Phys.Lett. {\bf 67}, 3924 (1995)


\bibitem{FujTar97}
T.Fujisawa and S.Tarucha,
Supperlatt. and Microstr. {\bf 21}, 247 (1997),
Jpn.J.Appl.Phys. {\bf 36}, 4000 (1997)

\bibitem{Oos97}
T.H.Oosterkamp et al,
Phys.Rev.Lett. {\bf 78}, 1536 (1997)


\bibitem{DatAna}
S.Datta and M.P.Anantram,
Phys.Rev. {\bf B45}, 13761 (1992)

\bibitem{Gri}
A.Grincwajg et al,
Phys.Rev. {\bf B52}, 12168 (1995)

\bibitem{Pas82}
A.B.Pashkovskii,
JETP {\bf 82}, 959 (1996)

\bibitem{ChuTan}
C.S.Chu and C.S.Tang,
Solid State Comm. {\bf 97}, 119 (1996)

\bibitem{YeyFlo91}
A.Levy Yeyati and F.Flores, 
Phys.Rev. {\bf B44},9020 (1991)


\bibitem{IvLev}
D.A.Ivanov and L.S.Levitov,
JETP Lett. {\bf 58}, 461 (1993)

\bibitem{LesLev}
G.B.Lesovik and L.S.Levitov,
Phys.Rev.Lett. {\bf 72}, 538 (1994)

\bibitem{Bor97}
V.S.Borovikov et al,
Fiz.Nizk.Temp.{\bf 23},313 (1997),
[see Low.Temp.Phys.{\bf 23}, 230 (1997)]

\bibitem{Lan92}
R.Landauer,
Physica Scripta,{\bf T42}, 110 (1992)



\bibitem{MF}
P.M.Morse and H.Feshbach, Methods of Theoretical Physics,
NY,Toronto, London, McGraw-Hill,1953
(chap.7)

\end{references}
\end{document}